\RequirePackage{fix-cm}

\documentclass[twocolumn,epjc3]{svjour3}  
\smartqed  
\RequirePackage{graphicx}

\RequirePackage{float}
\RequirePackage{amsmath}
\RequirePackage{amssymb}
\RequirePackage{lineno}
\RequirePackage[colorlinks,citecolor=blue,urlcolor=blue,linkcolor=blue]{hyperref}
\journalname{Preprint}

\modulolinenumbers[5] 

\begin{document}

\title{Unraveling the puzzle of slow components in gaseous argon of two-phase detectors for dark matter searches using Thick Gas Electron Multiplier
	}


\author{A.~Buzulutskov\thanksref{addr1,addr2}
        \and 
        E.~Frolov\thanksref{addr1,addr2,e1}
		\and
		E.~Borisova\thanksref{addr1,addr2}
		\and
		V.~Nosov\thanksref{addr1,addr2}
		\and
		V.~Oleynikov\thanksref{addr1,addr2}
		\and
		A.~Sokolov\thanksref{addr1,addr2}
        }

\thankstext{e1}{geffdroid@gmail.com (corresponding author)}


\institute{Budker Institute of Nuclear Physics SB RAS, Lavrentiev avenue 11, 630090 Novosibirsk, Russia \label{addr1} 
\and Novosibirsk State University, Pirogova street 2, 630090 Novosibirsk, Russia \label{addr2}
}

\date{Received: date / Accepted: date}

\maketitle

\begin{abstract}
The effect of proportional electroluminescence (EL) is used to record the primary ionization signal (S2) in the gas phase of two-phase argon detectors for dark matter particle (WIMP) searches and low-energy neutrino experiments. Our previous studies of EL time properties revealed the presence of two unusual slow components in S2 signal of two-phase argon detector, with time constants of about 4-5~$\mu$s and 50~$\mu$s. The puzzle of slow components is that their time constants and contributions to the overall signal increase with electric field (starting from a certain threshold), which cannot be explained by any of the known mechanisms of photon and electron emission in two-phase media. There are indications that these slow components result from delayed electrons, temporarily trapped during their drift in the EL gap on metastable negative argon ions of yet unknown nature. In this work, this hypothesis is confirmed by studying the time properties of electroluminescence in a Thick Gas Electron Multiplier (THGEM) coupled to the EL gap of two-phase argon detector. In particular, an unusual slow component in EL signal, similar to that observed in the EL gap, was observed in THGEM itself. In addition, with the help of THGEM operated in electron multiplication mode, the slow component was observed directly in the charge signal, confirming the effect of trapped electrons in S2 signal. These results will help to unravel the puzzle of slow components in two-phase argon detectors and thus to understand the background in low-mass WIMP searches.

\end{abstract}


  \section{Introduction}\label{intro}

Two-phase argon and xenon detectors for dark matter search experiments~\cite{Akimov21} have achieved the current best limits on WIMP-nucleon spin-independent cross-section~\cite{Aprile18,Aprile19,Agnes18b,Agnes18,LUX16}. These detectors measure both prompt primary scintillation signal (S1) and delayed primary ionization signal (S2), the latter being recorded in the gas phase using the effect of proportional electroluminescence (EL)~\cite{Buzulutskov20}. 

According to modern concepts~\cite{Buzulutskov20}, there are three mechanisms responsible for proportional EL in noble gases: excimer emission in the vacuum ultraviolet (VUV), emission due to atomic transitions in the near infrared (NIR), and neutral bremsstrahlung (NBrS) emission in the UV, visible and NIR range. These three mechanisms are referred to as excimer (ordinary) EL, atomic EL and NBrS EL, respectively. Let us briefly recall the details of these mechanisms using the example of Ar.

NBrS EL is due to bremsstrahlung of drifting electrons elastically scattered on neutral atoms \cite{Buzulutskov18,Borisova21,Borisova22,Aoyama22,Henriques22,Milstein22,Milstein23}:
\begin{eqnarray}
\label{Rea-NBrS-el}
e^- + \mathrm{Ar} \rightarrow e^- + \mathrm{Ar} + h\nu \; . 
\end{eqnarray}
It is fast ($\lesssim$1 ps) and has no threshold in energy and thus in electric field. 

Excimer EL is due to emission of noble gas excimers, in a singlet (Ar$^{*}_{2}(^{1}\Sigma^{+}_{u})$) or triplet (Ar$^{*}_{2}(^{3}\Sigma^{+}_{u})$) state, produced in three-body atomic collisions of the lowest excited atomic states, of Ar$^*(3p^54s)$ configuration, which in turn are produced by drifting electrons in electron-atom collisions (see reviews~\cite{Akimov21,Buzulutskov20,Chepel13,Buzulutskov17,Oliveira11}):
\begin{eqnarray}
\label{Rea-ord-el}
e^- + \mathrm{Ar} \rightarrow e^- + \mathrm{Ar}^{\ast}(3p^54s) \; , \nonumber \\
\mathrm{Ar}^{\ast}(3p^54s) + 2\mathrm{Ar} \rightarrow \mathrm{Ar}^{\ast}_2(^{1,3}\Sigma^{+}_{u}) + \mathrm{Ar} \; , \nonumber \\
\mathrm{Ar}^{\ast}_2(^{1,3}\Sigma^{+}_{u}) \rightarrow 2\mathrm{Ar} + h\nu \; .
\end{eqnarray}
It has a threshold in reduced electric field (E/N), of about 4~Td (1~Td$=10^{-17}$~V~cm$^2$) \cite{Oliveira13,Buzulutskov22}, corresponding to excitation of the lower energy levels of Ar$^{*}(3p^{5}4s)$ configuration. Singlet and triplet excimer states are responsible for respectively the fast (4.2~ns) and slow (3.1~$\mu$s) component of excimer EL \cite{Buzulutskov17}.

Atomic EL is due to atomic transitions between the higher (Ar$^*(3p^54p)$) and lower (Ar$^*(3p^54s)$) excited states, the former being also produced by drifting electrons~\cite{Oliveira13,Buzulutskov11}:
\begin{eqnarray}
e^- + \mathrm{Ar} \rightarrow e^- + \mathrm{Ar}^{\ast}(3p^54p) \; , \nonumber \\
\mathrm{Ar}^*(3p^54p) \rightarrow \mathrm{Ar}^*(3p^54s)+h\nu \; .
\end{eqnarray}
It is fast (20-40~ns) and has a threshold in reduced electric field of about 5~Td \cite{Oliveira13,Buzulutskov22}, corresponding to excitation of the higher energy levels of Ar$^{*}(3p^{5}4p)$ configuration.

Depending on the reduced electric field, some of various EL mechanisms dominate over the others, as illustrated in~\cite{Buzulutskov20}. Namely, NBrS EL fully dominates below 4~Td. Excimer EL dominates above 4~Td in terms of the absolute photon yield in the VUV range. However, in the visible and NIR range, NBrS EL remains the main mechanism up to 10~Td, while above 10~Td it is taken over by atomic EL. 

At low WIMP masses, near the detector threshold, only the S2 signal is possible to detect, using the so-called S2-only analysis~\cite{Aprile19,Agnes18}. In this case, a precise model of expected background rates for S2 signals as a function of detected number of electrons is required. At the moment, there is discrepancy between the observed and expected number of events in low-energy region, at low number of electrons, in DarkSide-50~\cite{Agnes18} and XENON1T~\cite{Aprile20} experiments. Such discrepancy is currently interpreted as being due to electrons trapped on impurities in the liquid bulk or at the liquid-gas interface both in Xe~\cite{Aprile22} and Ar~\cite{Agnes23}. Earlier observations of slow components and delayed pulses on $\mu$s and ms scales in Xe \cite{Aprile14,Akimov16,Sorensen18,Tomas18,Akerib20,Kopec21} indicate the presence of these delayed electrons in two-phase Xe detectors.

\begin{figure}[!t]
\centering
\includegraphics[width=1.0\linewidth]{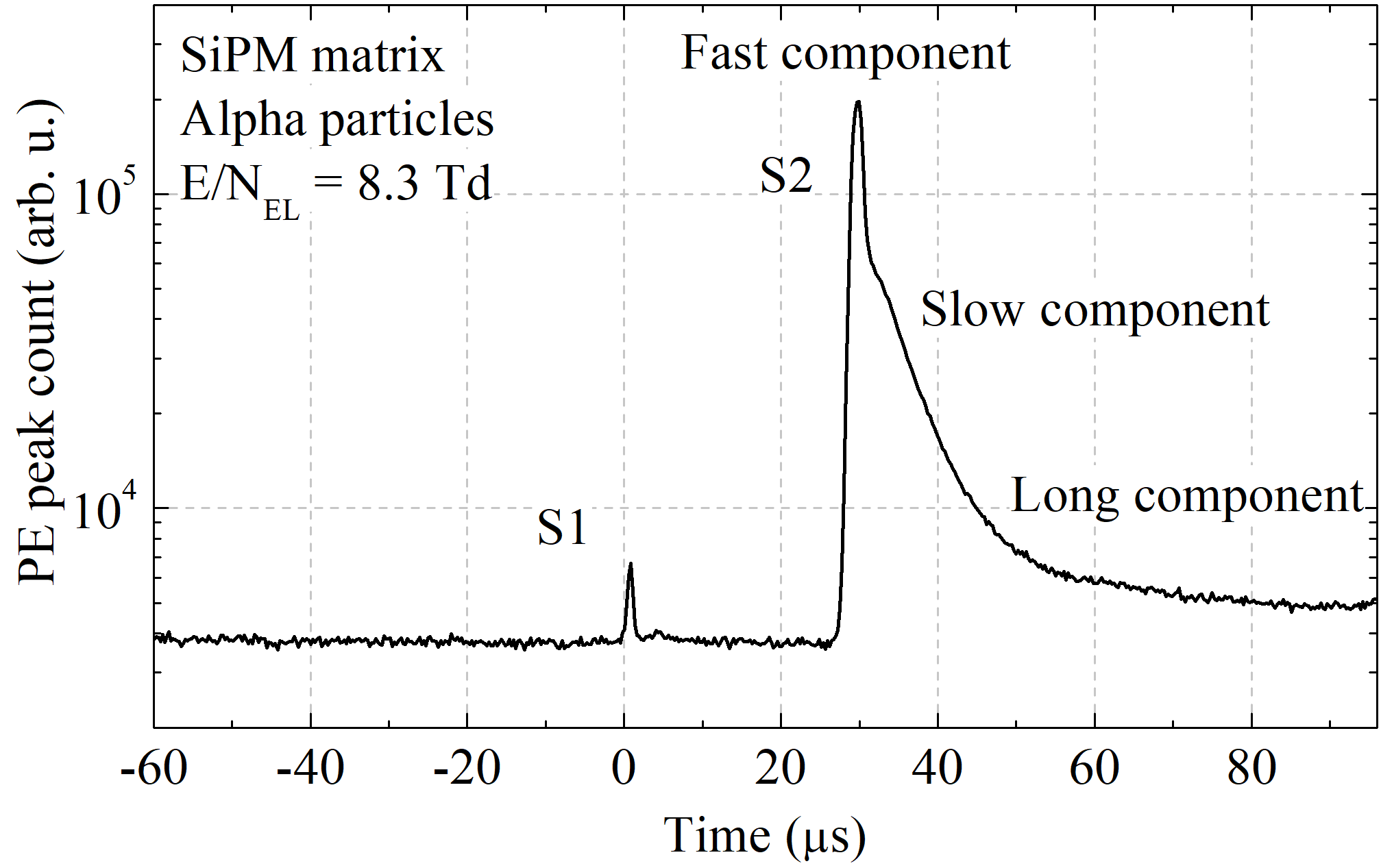}
\caption{Signal pulse shape from SiPM matrix facing the EL gap of two-phase Ar detector obtained in~\cite{Buzulutskov22} due to NBrS EL in the visible and NIR range, with $^{238}$Pu alpha particles at reduced electric field of 8.3~Td and pressure of 1.00~atm. One can see the S1 signal as well as the S2 signal with one fast and two slow components.}
\vspace{-10pt}
\label{fig:S1_S2_shape}
\end{figure}

In our previous works~\cite{Buzulutskov22,Bondar20,Bondar20b} it was shown that there are likely delayed electrons of another nature in two-phase Ar detectors. In those works, the S2 time properties were studied in a wide range of electric fields using proportional EL both in the VUV range and in that of visible and NIR. The former is provided by excimer EL while the latter is provided by NBrS EL. Two unusual slow components with time constants of about 4-5~$\mu$s and 50~$\mu$s, referred to as ``slow'' and ``long'' component respectively, were observed. An example pulse shape showing these components obtained in~\cite{Buzulutskov22} at the highest electric field and in the visible range is shown in Fig.~\ref{fig:S1_S2_shape}. 

These slow components have the following puzzling properties that have never been observed before for slow components of conventional (excimer) scintillation:

1. There are two slow components in proportional EL, observed simultaneously in the VUV and in the visible and NIR range, with different time constants: of about 4-5~$\mu$s and 50~$\mu$s.

2. Both slow components emerge at a certain threshold in reduced electric field, of about 5~Td, regardless of the gas phase density, which is 1~Td above the onset of excimer EL, the latter being related to the lower excited atomic states Ar$^{*}(3p^{5}4s)$. Accordingly, the 5 Td threshold is related to the higher atomic excited states Ar$^{*}(3p^{5}4p)$, similarly to atomic EL in the NIR which has the same threshold, of 5~Td (see discussion in \cite{Buzulutskov22}).

3. Another puzzling property of slow components is that their contributions and time constants increase with electric field, which cannot be explained by either of the two known mechanisms of slow component formation in two-phase Ar detectors, namely by electron emission from liquid to gas phase~\cite{Borghesani90,Bondar09} and by VUV photon emission via triplet excimer state Ar$_{2}^{*}$($^{3}\Sigma_{u}^{+}$)~\cite{Akimov21,Buzulutskov17}. In those mechanisms, the slow component contribution and time constant either decrease with electric field or do not depend on the electric field at all, respectively. 

4. An unexpected temperature dependence of the 50~$\mu$s component (and presumably that of 5~$\mu$s) was observed in gaseous Ar: its contribution decreased with temperature, practically disappearing at room temperature. 

All puzzling properties of slow components can be successfully explained in the framework of hypothesis \cite{Buzulutskov22} that these are produced in the EL gap in the charge signal itself, due to temporarily trapping (attachment) of drifting electrons on metastable negative Ar ions of yet unknown nature with lifetimes close to time constants of the slow components, i.e. about 4 and 50~$\mu$s. Taking into the account the 5 Td threshold, the formation of these metastable Ar ions is related either to the Ar$^{*}(3p^{5}4p)$ states or some unknown state with similar energy level. In the frame of this approach it is possible to explain the temperature dependence of the slow components, if one assumes that they are produced mostly at low temperatures due to electron collisions with Van-der-Waals molecules $\mathrm{Ar}_2(\mathrm{X}\, ^1\Sigma^{+}_{g})$~\cite{Buzulutskov17,Smirnov84,Smirnov96,Stogrin59}. 

In this work, this hypothesis is verified by studying the time properties of electroluminescence in a Thick Gas Electron Multiplier (THGEM) coupled to the EL gap of two-phase argon detector. The THGEM thus worked as an additional ``EL gap'' of smaller thickness, decoupled from the liquid-gas interface. If the hypothesis is correct, the slow components should be observed both in the EL gap and THGEM. In addition, with the help of THGEM operated in an electron multiplication (avalanche) mode, it is possible to observe the slow component directly in the charge signal itself, confirming the effect of trapped electrons in S2 signal. These studies will help to unravel the puzzle of slow components in two-phase Ar detectors and thus to understand the background in low-mass WIMP searches.

\section{Experimental setup}\label{SetupSection}
\begin{figure}[!t]
\centering
\includegraphics[width=1.0\linewidth]{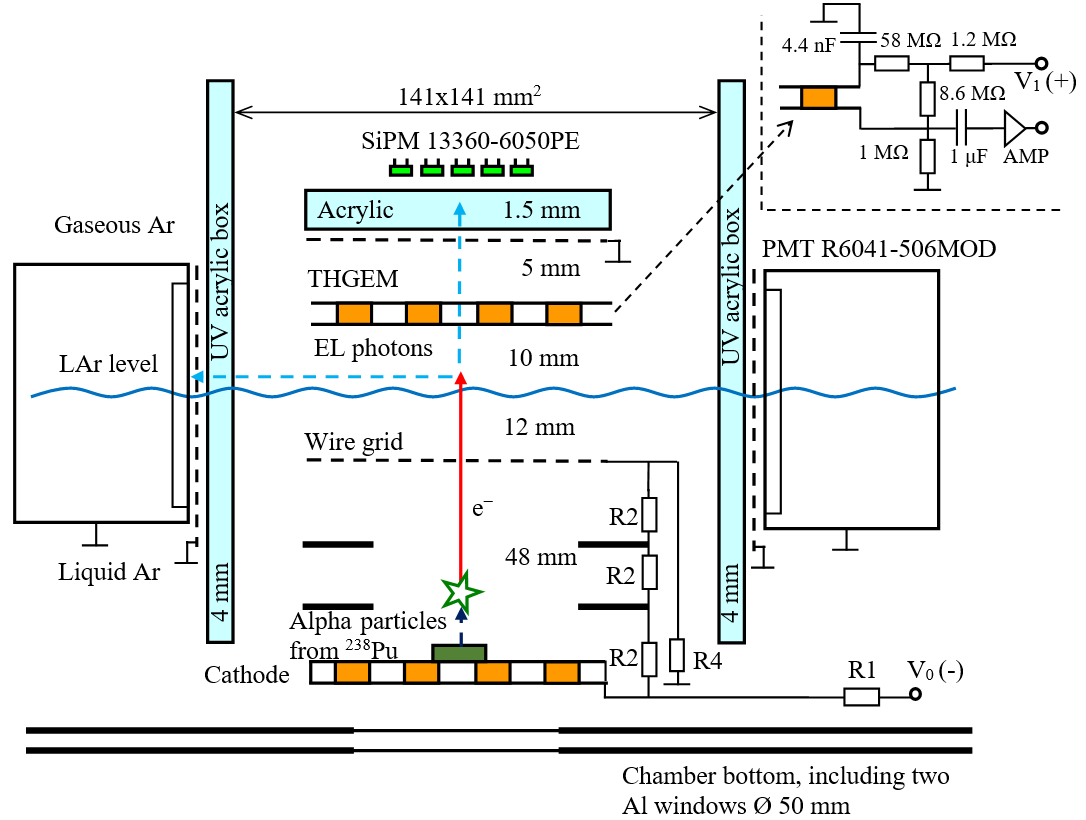}
\caption{Schematic view of the experimental setup (not to scale).}
\vspace{-10pt}
\label{fig:setup_scheme}
\end{figure}

Fig.~\ref{fig:setup_scheme} shows the scheme of the experimental setup. It included 48~mm drift region (from the cathode to the wire grid), 12~mm electron emission region (from the wire grid to the liquid-gas interface) and 10~mm EL gap (from the liquid-gas interface to the THGEM). Electric field in these regions was defined by resistors R1~=~120~M$\Omega$, R2~=~40~M$\Omega$ and R4~=~600~M$\Omega$, resistors R2 being connected to field shaping rings necessary to obtain uniform field in the drift region~\cite{Bondar19}.

Compared to our previous work~\cite{Buzulutskov22}, there are only a few modifications: interface THGEM (THGEM0) was replaced by the wire grid with corresponding R3 resistor being removed; THGEM above the EL gap had an active area of 10$\times$10~cm$^2$, dielectric thickness of 0.4~mm, hole pitch of 0.9~mm, rim of 0.1~mm and hole diameter of 0.5~mm (28\% optical transparency at normal incidence). Also, THGEM was connected to a voltage divider with charge readout circuit, as shown in Fig.~\ref{fig:setup_scheme}. The detector operated at saturated argon pressures of 1.00 and 1.50 atm, corresponding to a temperature of 87.3 and 91.3~K and gas density of 8.71$\times$10$^{19}$ and 1.26$\times$10$^{20}$ cm$^{-3}$ respectively~\cite{Stewart89,Fastovsky72}. 

\sloppy Of crucial importance for performance of two-phase detectors is Ar purity. In order to purify it from electronegative impurities, Ar gas was liquefied from a storage bottle into the cryogenic chamber while passing through Oxysorb filter at the start of each experimental run. At the end of the run, Ar was collected from the chamber back into the bottle. Thus, the same Ar gas was purified multiple times from its initial total impurity content below 2~ppm (as declared by manufacturer). This filtration process was demonstrated to achieve electron life-time in liquid Ar $>$~100~$\mu$s at 200~V/cm field~\cite{Bondar17}, which corresponds to oxygen content below 3~ppb~\cite{Li22}. Additionally, the N$_2$ content was monitored by a gas analyzer SVET~\cite{SVET} based on an emission spectrum measurement technique: it was below 1~ppm.

The EL signal was recorded using four PMTs R6041-506MOD~\cite{Bondar15,Bondar17a} and 5$\times$5 SiPM matrix composed of 13360-6050PE type SiPMs~\cite{Hamamatsu} as shown in Fig.~\ref{fig:setup_scheme}. Since no WLS was used, the devices were sensitive to EL in the visible and NIR range only, provided by NBrS and atomic EL (see section~\ref{intro}). The recorded photoelectrons (PE) from selected events were used to produce time histograms reflecting the pulse shape averaged over the selected events. This pulse shape was then used to obtain time constants and contributions of slow components if any are present. A detailed description of the analysis algorithm, event selection and determination of the parameters of slow components can be found elsewhere~\cite{Buzulutskov22,Bondar20}.

The charge readout circuit (denoted as AMP in Fig.~\ref{fig:setup_scheme}) consisted of CAEN MOD A1422 charge amplifier with 45~mV/MeV gain 
and ORTEC 570 research amplifier with 0.5~$\mu$s shaping time and gain of 20 connected in series. The amplified signal was split using CAEN N625 unit and sent to oscilloscope and CAEN V1720E flash ADC for recording on PC for off-line analysis.

The measurements were performed using 5.5~MeV alpha particles from a $^{238}$Pu source placed on a cathode immersed in liquid Ar. This source provided rather localized ionization along the drift direction, suitable for studying the time properties of both EL and charge signals. The trigger was taken from the S2 signal, provided by the sum of the signals from all four PMTs. More details on the experimental setup and procedures can be found in~\cite{Buzulutskov22,Bondar20}.

\section{THGEM performance}\label{THGEM}

THGEM coupled to the EL gap is the main feature of this study. As shown in Fig.~\ref{fig:setup_scheme}, it was connected to a voltage divider which defined two modes of its operation: passive, when the divider voltage was zero and THGEM acted as just a grounded anode of the EL gap, and active, when the divider voltage was positive and THGEM acted as the second EL gap with an order of magnitude smaller thickness. In the first case there was only one EL gap, similarly to our previous works which allowed to reproduce and verify the previous results. In the second case there were two successive EL gaps, THGEM being the second gap. Moreover, THGEM was operated not only in proportional EL mode (when light amplitude increases proportionally with the electric field), but also in avalanche (charge multiplication) EL mode when light and charge amplitude increases exponentially with the electric field. Maximum voltage across THGEM (and maximum amplification) was limited by discharges which occur when electron avalanches become self-sustainable at sufficiently high voltages. 

In this work, we refer to the signal induced on THGEM electrodes by drifting electrons as simply ``charge signal''. Using THGEM allowed us to study charge signal not only though EL, but also directly with charge-sensitive amplifier. Due to poor signal-to-noise ratio, the charge signal produced by alpha particles from $^{238}$Pu source (of about 4000 e$^-$~\cite{Hitachi87,Masuda89}) could not be detected directly in the EL gap when THGEM was grounded and operated as an anode. Accordingly, to reliably detect the charge signal produced by alpha particles, THGEM was operated in an avalanche mode, by applying the voltage (V$_1$) to its divider, as shown in Fig.~\ref{fig:setup_scheme}.

\begin{figure}[!t]
\centering
\includegraphics[width=1.0\linewidth]{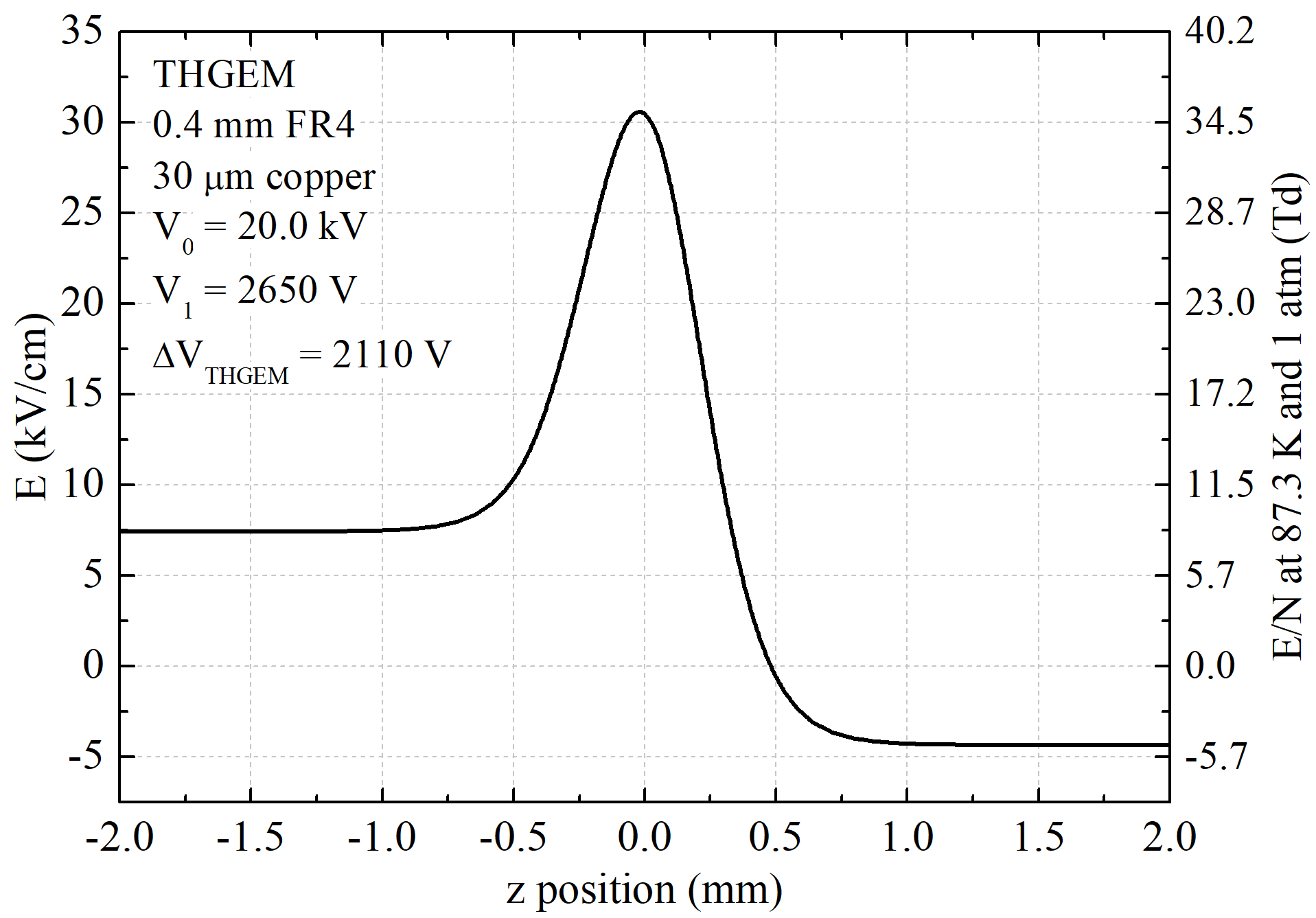}
\caption{Calculated electric field along the vertical axis going through the THGEM hole center.}
\vspace{-10pt}
\label{fig:THGEM_field}
\end{figure}

Many results in this work are shown as a function of reduced electric field at the center of the THGEM hole (E/N$_{\textrm{THGEM}}$). Electric field map in THGEM was obtained by calculating electrostatic potentials numerically using finite elements method with Gmsh~\cite{Gmsh,Geuzaine2009} and Elmer~\cite{Elmer} open-source programs. These calculations are discussed in detail in~\cite{Bondar19}. Fig.~\ref{fig:THGEM_field} shows the electric field along the vertical axis passing trough the THGEM hole center calculated at maximal V$_0$ and V$_1$ used in this work. Note that the field at the THGEM hole center is equal to 0.57 of its ``nominal'' value, the latter defined as the voltage applied across THGEM (THGEM voltage, $\Delta$V$_{\textrm{THGEM}}$) divided by the dielectric thickness.
It should be noted that in the present work the systematic uncertainty of reduced electric field both in the EL gap and at the THGEM hole center is estimated to be 4\%.


\begin{figure}[!t]
\centering
\includegraphics[width=1.0\linewidth]{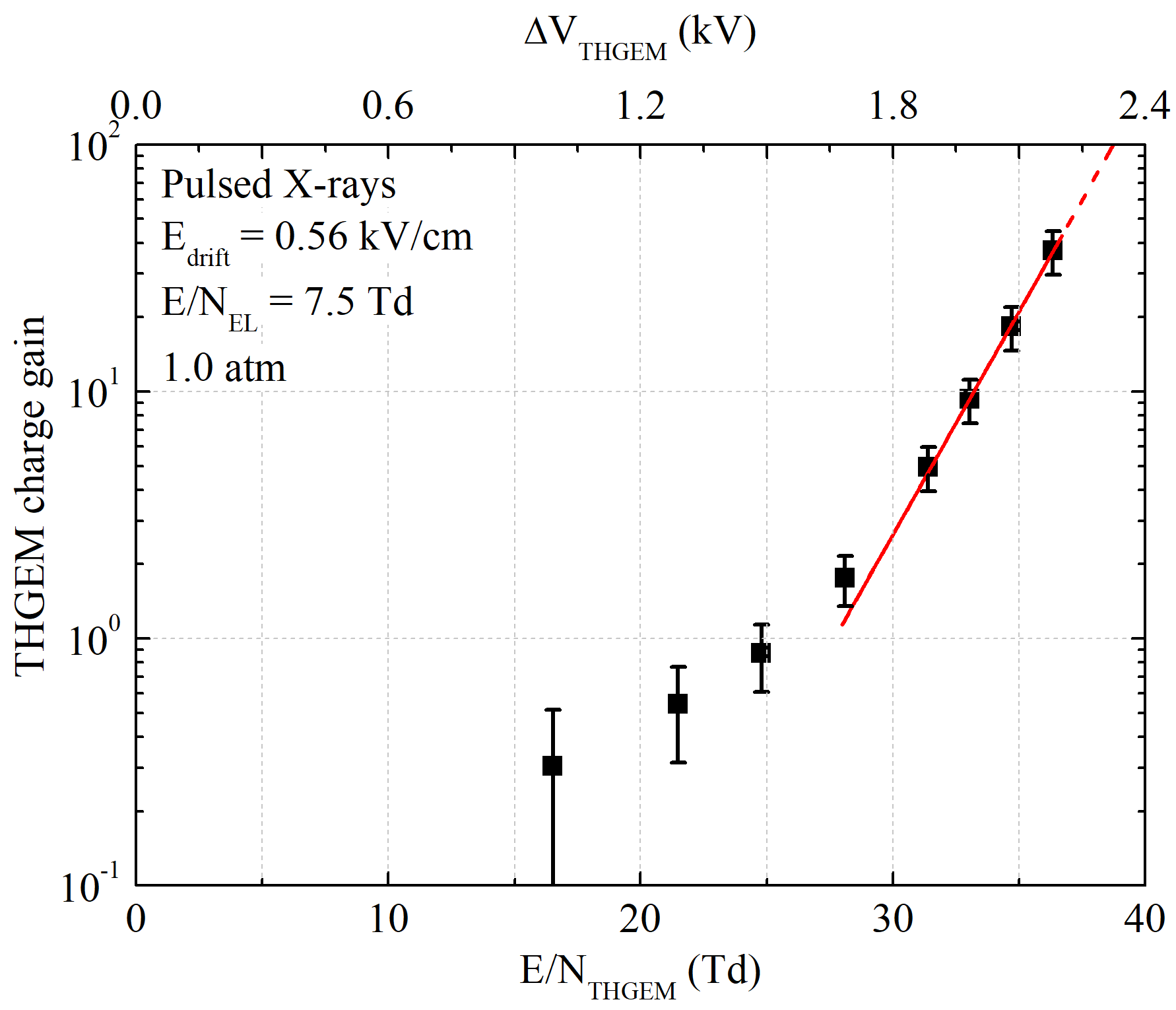}
\caption{Effective charge gain of THGEM used in this work as a function of reduced electric field at its hole center, measured in two-phase Ar detector at 1.0~atm and 87.3~K~\cite{Aalseth21}. Top axis shows corresponding voltage across it. The gain was measured at high drift (0.56 kV/cm) and EL (7.5 Td) fields using pulsed X-ray tube.}
\vspace{-10pt}
\label{fig:THGEM_gain}
\end{figure}

\begin{figure}[!t]
\centering
\includegraphics[width=1.0\linewidth]{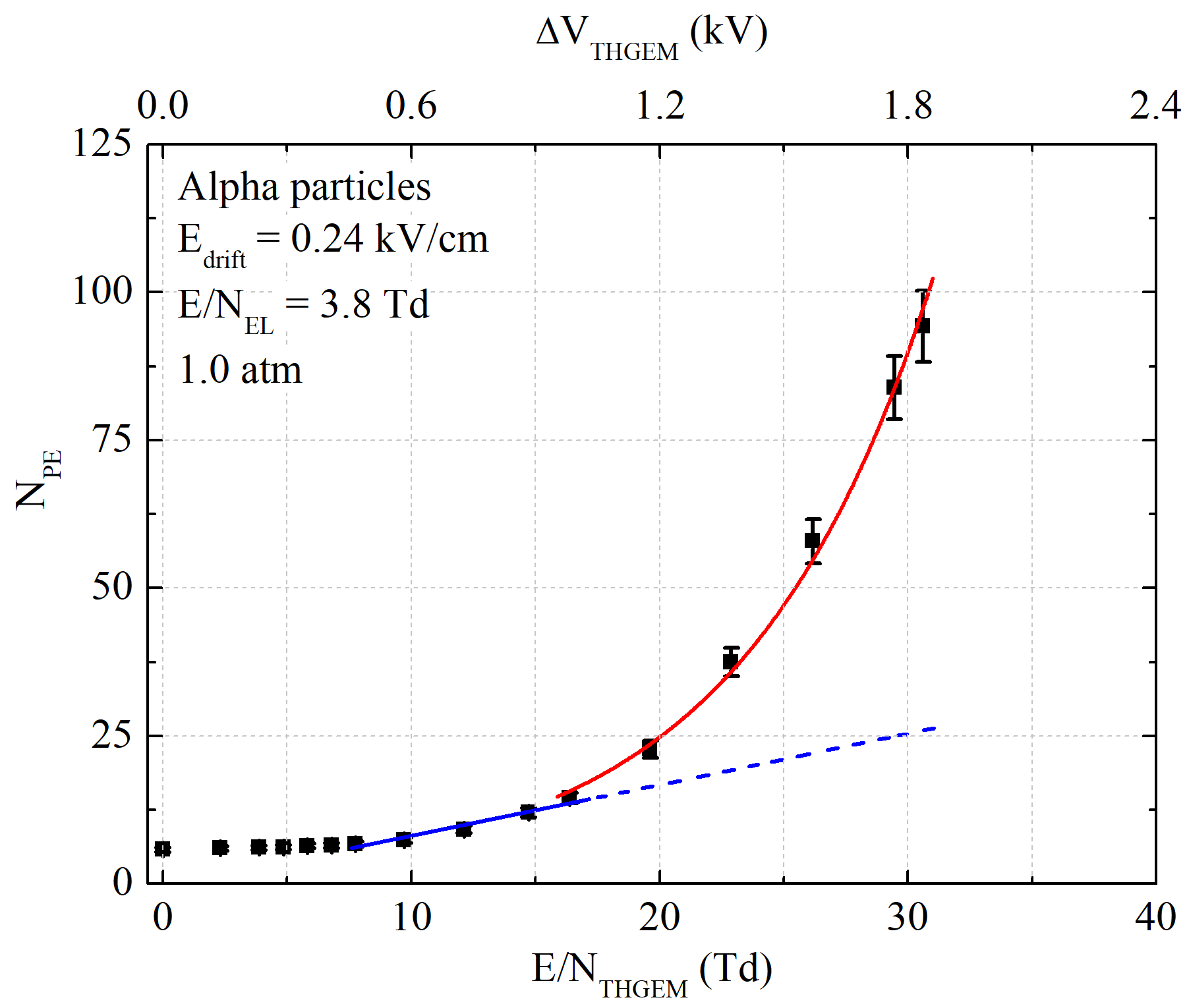}
\caption{SiPM-matrix signal amplitude, expressed in the number of recorded photoelectrons, as a function of reduced electric field at the THGEM hole center. The data were obtained in two-phase Ar detector at 1.0~atm and 87.3~K at low drift and EL fields using $^{238}$Pu alpha particles with EL gap thickness of 10 mm. The blue and red lines are respectively the linear and exponential fits to the data points and correspond to the proportional EL and avalanche mode of THGEM performance.
}
\vspace{-10pt}
\label{fig:THGEM_Npe}
\end{figure}

Fig.~\ref{fig:THGEM_gain} shows the effective THGEM gain in gaseous Ar at 1.0~atm as a function of the voltage across it ($\Delta$V$_{\textrm{THGEM}}$), measured in two-phase detector at high electric field in the EL gap (7.5 Td). These measurements were conducted in~\cite{Aalseth21} with high-intensity pulsed X-ray tube providing about 10$^5$ electrons that reach the gas phase. The effective THGEM gain is defined as the charge recorded on the THGEM electrode divided by the charge in front of THGEM, i.e. by charge drifting in the EL gap. The latter was measured with grounded THGEM. It should be noted that the effective gain is always smaller than the real gain of THGEM, proportional to the real number of avalanche electrons~\cite{Bellazzini98}, since a some part of electrons in front of THGEM do not enter its holes and thus are lost for multiplication, as part of the electric field lines converge at the bottom electrode. 

Another way to characterize the THGEM performance is to use an optical signal of the SIPM matrix placed behind THGEM (see Fig.~\ref{fig:setup_scheme}). Its amplitude, being expressed in the number of photoelectrons, recorded mostly due to atomic EL in the NIR~\cite{Buzulutskov20}, is roughly proportional to the number of avalanche electrons in the THGEM holes and thus to the real THGEM gain, see Fig.~\ref{fig:THGEM_Npe}.

As the reduced electric field increases, one can observe three modes of THGEM performance in Fig.~\ref{fig:THGEM_Npe}: that of weak NBrS EL in the EL gap (below 7 Td), that of proportional EL in THGEM due to atomic EL in the NIR (between 7 and 17 Td) described by a linear function (blue line), and that of avalanche scintillation EL in THGEM due to atomic EL in the NIR (above 17 Td) described by an exponential function (red line). The electric field at which transition from proportional to an exponential mode occurs is in agreement with detailed measurements conducted elsewhere~\cite{Hollywood20}.

\section{Observation of unusual slow components in THGEM}\label{results_slow_comp}

\begin{figure}[!t]
\centering
\includegraphics[width=1.0\linewidth]{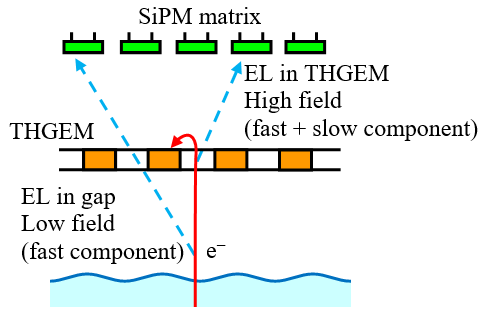}
\caption{Concept of experiment to observe the unusual slow components in THGEM itself.}
\vspace{-10pt}
\label{fig:THGEM_concept}
\end{figure}

\begin{figure}[!t]
\centering
\includegraphics[width=1.0\linewidth]{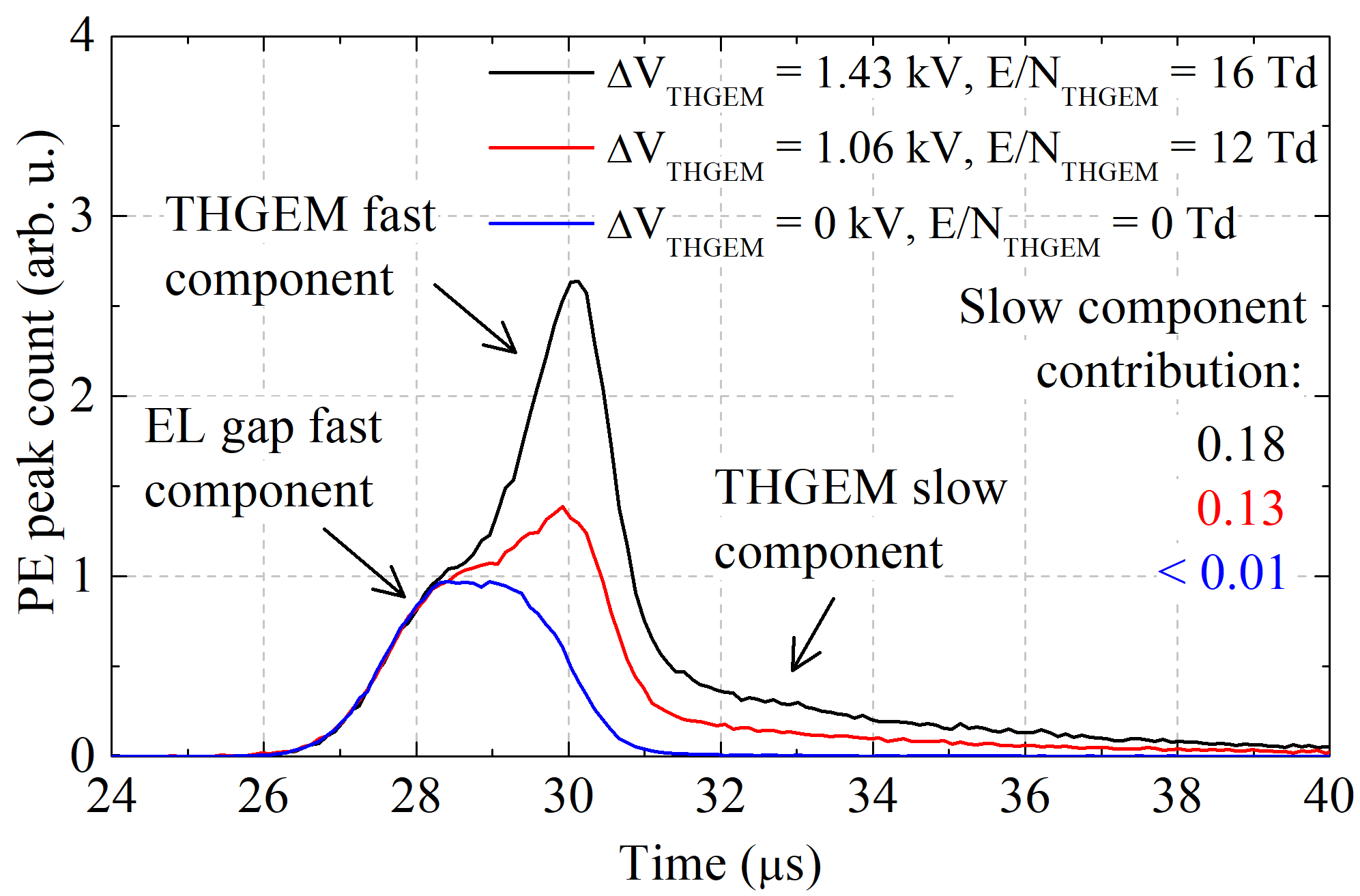}
\caption{Pulse shape of EL signal in two-phase Ar detector for SiPM-matrix readout at low EL gap field, of 3.7~Td, at different THGEM voltages corresponding to the reduced electric field in the THGEM hole center of 0, 11.9 and 16.1 Td. The data were obtained with $^{238}$Pu alpha particles and 10~mm EL gap thickness at 1.5~atm pressure. 
}
\vspace{-10pt}
\label{fig:THGEM_shapes}
\end{figure}

In our previous work~\cite{Buzulutskov22}, one of the main results was that the unusual slow components in the EL signal emerge at electric fields higher than 4.8~$\pm$~0.2~Td and that with increasing field both their time constants and contributions also increase. It was also shown that the unusual slow components are most likely formed in the gas phase in the EL gap. To verify these results, THGEM was used in this work as an additional gas gap, coupled to the EL gap and decoupled from the liquid-gas interface. 

To demonstrate that the unusual slow components can be produced in THGEM itself, the EL gap field was fixed at low value of 3.7~Td, just below the onsets of usual slow component due to triplet excimer EL and unusual slow components due to electron trapping. Consequently, one would expect here that only the fast component due to NBrS EL will be present in the EL gap signal at zero THGEM voltage. On the other hand, one would expect that with increasing the THGEM voltage the slow components will emerge when the reduced electric field in the THGEM holes exceeds 4.8~Td. The idea of such an experiment is schematically depicted in Fig.~\ref{fig:THGEM_concept}.

\sloppy Fig.~\ref{fig:THGEM_shapes} fully confirms these expectations: at zero THGEM voltage only the fast component due to NBrS EL in the EL gap (blue line) is observed, with almost symmetric shape defined by electron drift time through the EL gap, electron diffusion and trigger conditions. As $\Delta$V$_{\textrm{THGEM}}$ increases, EL from THGEM becomes noticeable and its own fast and slow components are added to the EL signal of the EL gap, with some delay. The fast component of THGEM, as expected, is delayed with respect to that of the EL gap by about 1~$\mu$s and thus can be separated from it. In addition, at high E/N$_{\textrm{THGEM}}$ values (red and black lines in Fig.~\ref{fig:THGEM_shapes}), the THGEM slow component becomes distinct, its time constant (of about 4 $\mu$s) and contribution increasing with the electric field. It should be noted that contribution of slow components is defined here in the same way as in~\cite{Bondar20b}: the total contribution of slow and long component was calculated as a fraction of pulse-shape area lying after certain time threshold. Slow and long component were separated from each other by exponential fits.

\begin{figure}[!t]
\centering
\includegraphics[width=1.0\linewidth]{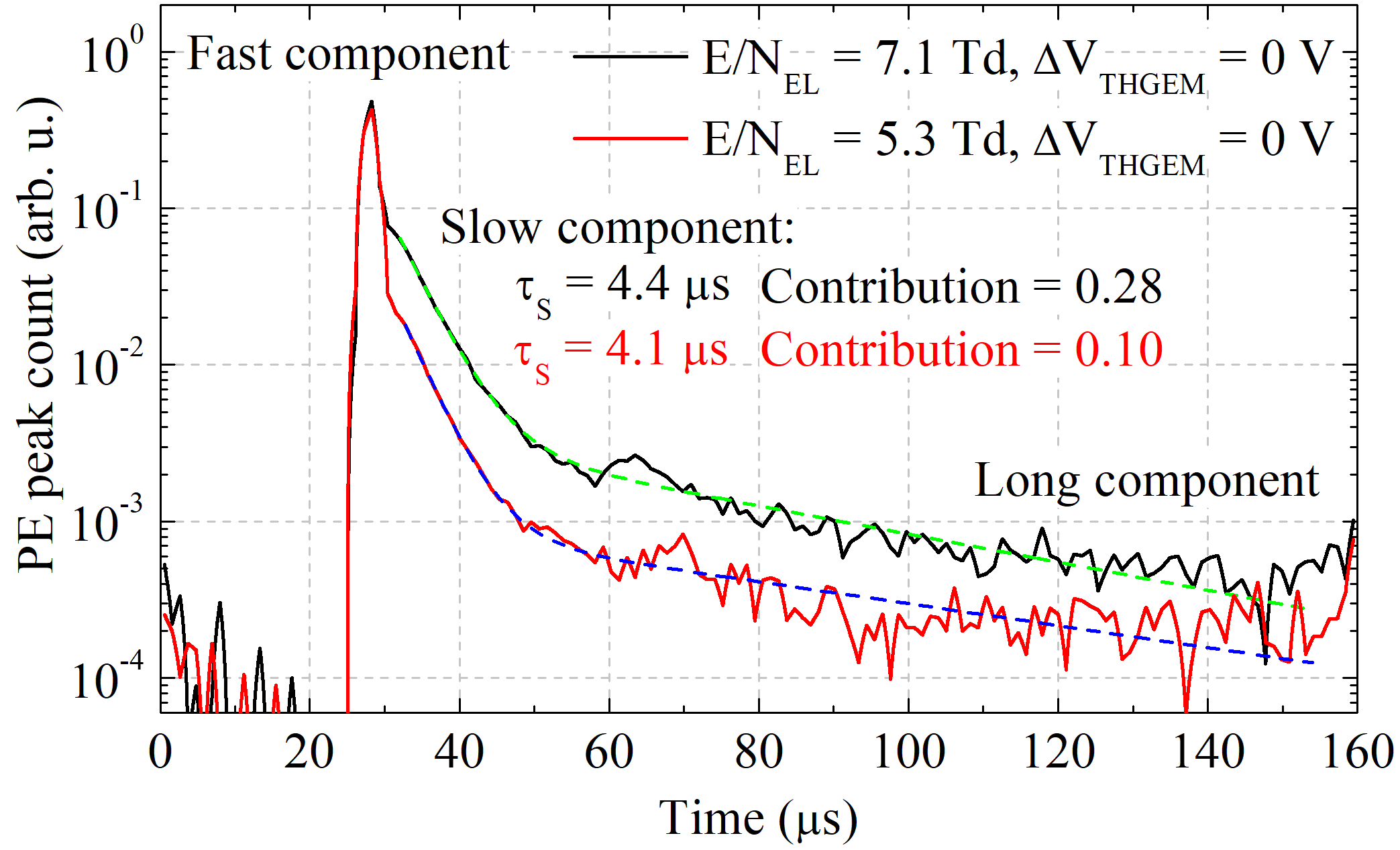}
\caption{Pulse shapes of EL signal in two-phase Ar detector for SiPM-matrix readout with grounded THGEM at two different reduced electric fields, of 5.3 and 7.1 Td. The data were obtained with $^{238}$Pu alpha particles and 10~mm EL gap thickness at 1.5~atm pressure. Dashed lines show fits of slow and long components by sum of two exponents (see~\cite{Bondar20b} for details).
}
\vspace{-10pt}
\label{fig:EL_gap_shapes_log}
\end{figure}

\begin{figure}[!t]
\centering
\includegraphics[width=1.0\linewidth]{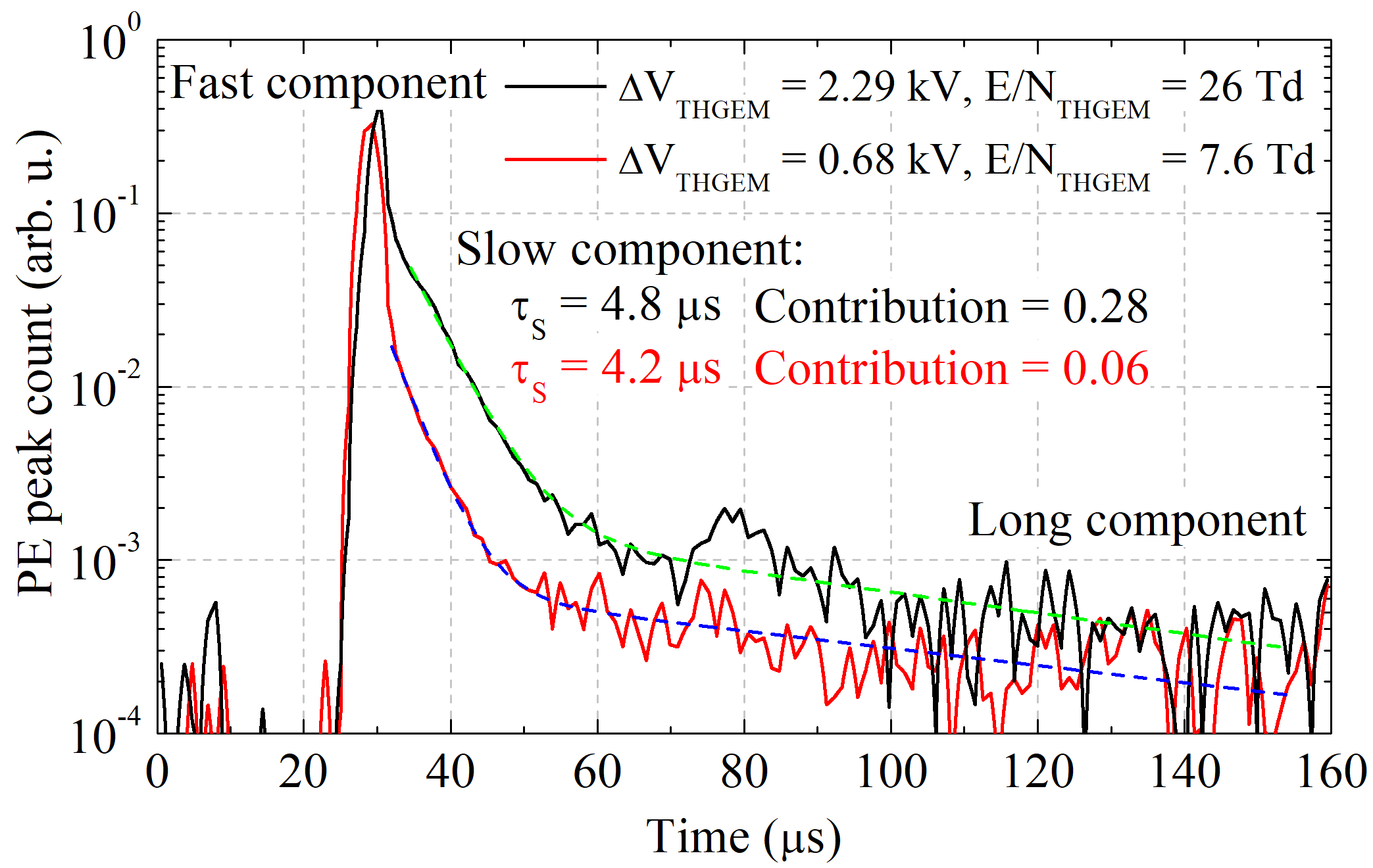}
\caption{Pulse shapes of EL signal in two-phase Ar detector for SiPM-matrix readout at low EL gap field, of 3.7~Td, at two different THGEM voltages corresponding to the reduced electric field in the THGEM hole center of 7.6 and 26 Td. The data were obtained with $^{238}$Pu alpha particles and 10~mm EL gap thickness at 1.5~atm pressure. Dashed lines show fits of slow and long components by sum of two exponents (see~\cite{Bondar20b} for details).
}
\vspace{-10pt}
\label{fig:THGEM_shapes_log}
\end{figure}

Fig.~\ref{fig:EL_gap_shapes_log} and~\ref{fig:THGEM_shapes_log} compare the slow components produced in the EL gap and in THGEM by corresponding pulse shapes. Fig.~\ref{fig:EL_gap_shapes_log} shows the EL signal pulse shapes when the EL gap had high electric field and THGEM was grounded (passive). Fig.~\ref{fig:THGEM_shapes_log} shows results obtained with active THGEM and the EL gap field being below the slow component threshold. In both cases, the fast, slow and long components are observed with similar time constants and contributions. These figures demonstrate the strong similarity between the unusual slow components produced in the EL gap and those produced in THGEM both in their general shape and in their dependency on the electric field. 

Compared to previous results, one can see additional so-called S3 signals that look like wide peaks in Fig.~\ref{fig:EL_gap_shapes_log} at around 62 and 70~$\mu$s and in Fig.~\ref{fig:THGEM_shapes_log} at around 80~$\mu$s. The time positions of the peaks change with the electric field, their delay with respect to fast component corresponding to electron drift time from the cathode. These are obviously caused by photon feedback from the EL region due to photo-ionization of the cathode, which became stronger when the interface electrode was replaced from THGEM plate (used in previous works) to wire grid in this work. 

Note that the shape of the slow component is more complex than just the shape of exponentially decreasing function: it tends to have a small bump at the beginning, in particular at higher electric fields, in accordance with our previous results on EL pulse shapes~\cite{Buzulutskov22,Bondar20b}. It is most probably due to photon feedback to the interface electrode (wire grid here and THGEM plate in previous works), similarly to S3 signal produced on the cathode. 

\begin{figure}[!t]
\centering
\includegraphics[width=1.0\linewidth]{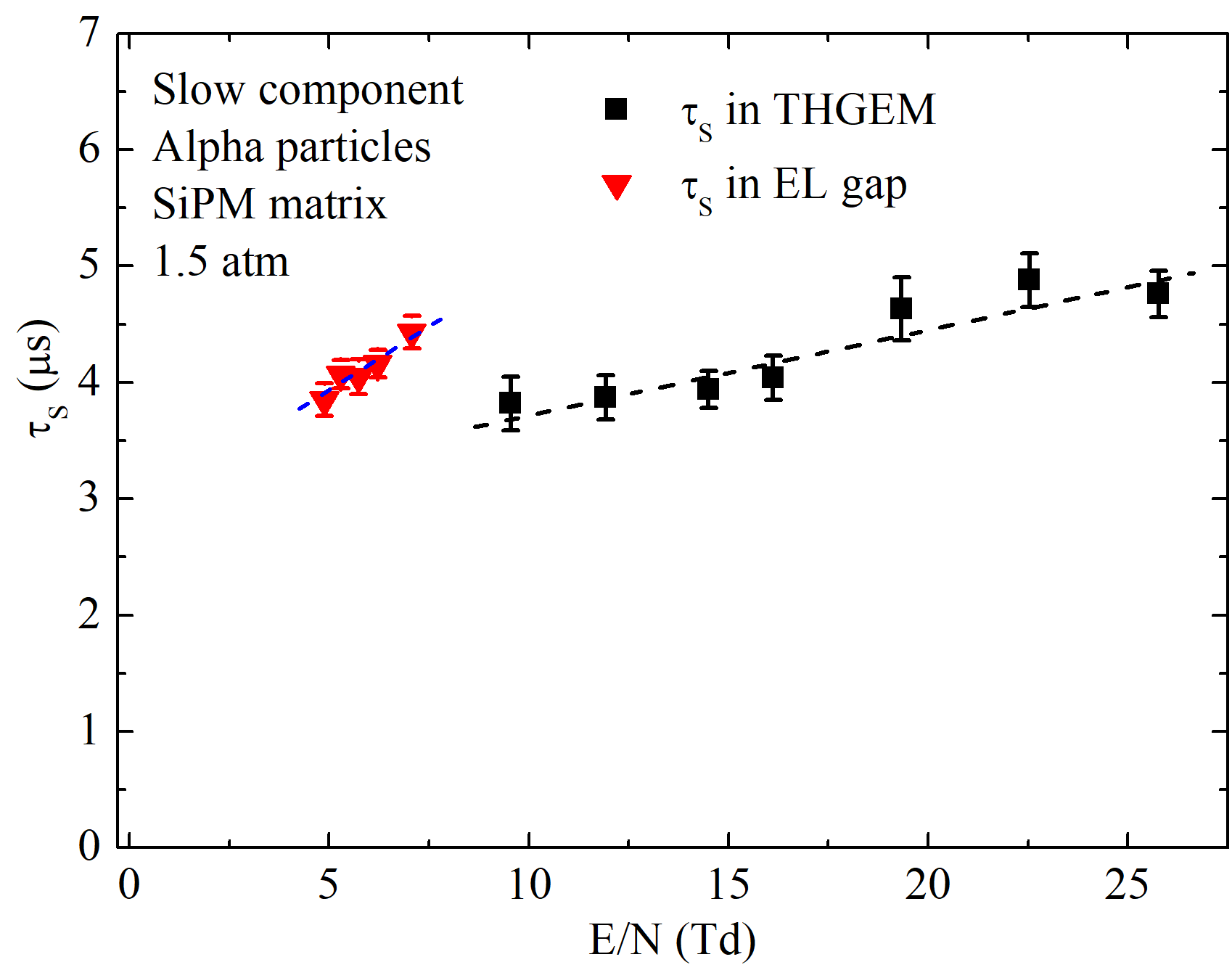}
\caption{Time constant ($\tau_{S}$) of the slow component of EL signal in THGEM and EL gap as a function of the reduced electric field in the THGEM hole center and in the EL gap respectively. The measurements in THGEM were conducted with active voltage divider shown in Fig.~\ref{fig:setup_scheme} at fixed low EL gap field, of 3.7~Td, while those in EL gap were done with grounded THGEM acting as an anode of the gap. The data were obtained in two-phase Ar detector with 10~mm thick EL gap and 1.5~atm pressure using $^{238}$Pu alpha particles. The dashed lines are linear fits to the data points.
}
\vspace{-10pt}
\label{fig:THGEM_vs_gap_tau}
\end{figure}

\begin{figure}[!t]
\centering
\includegraphics[width=1.0\linewidth]{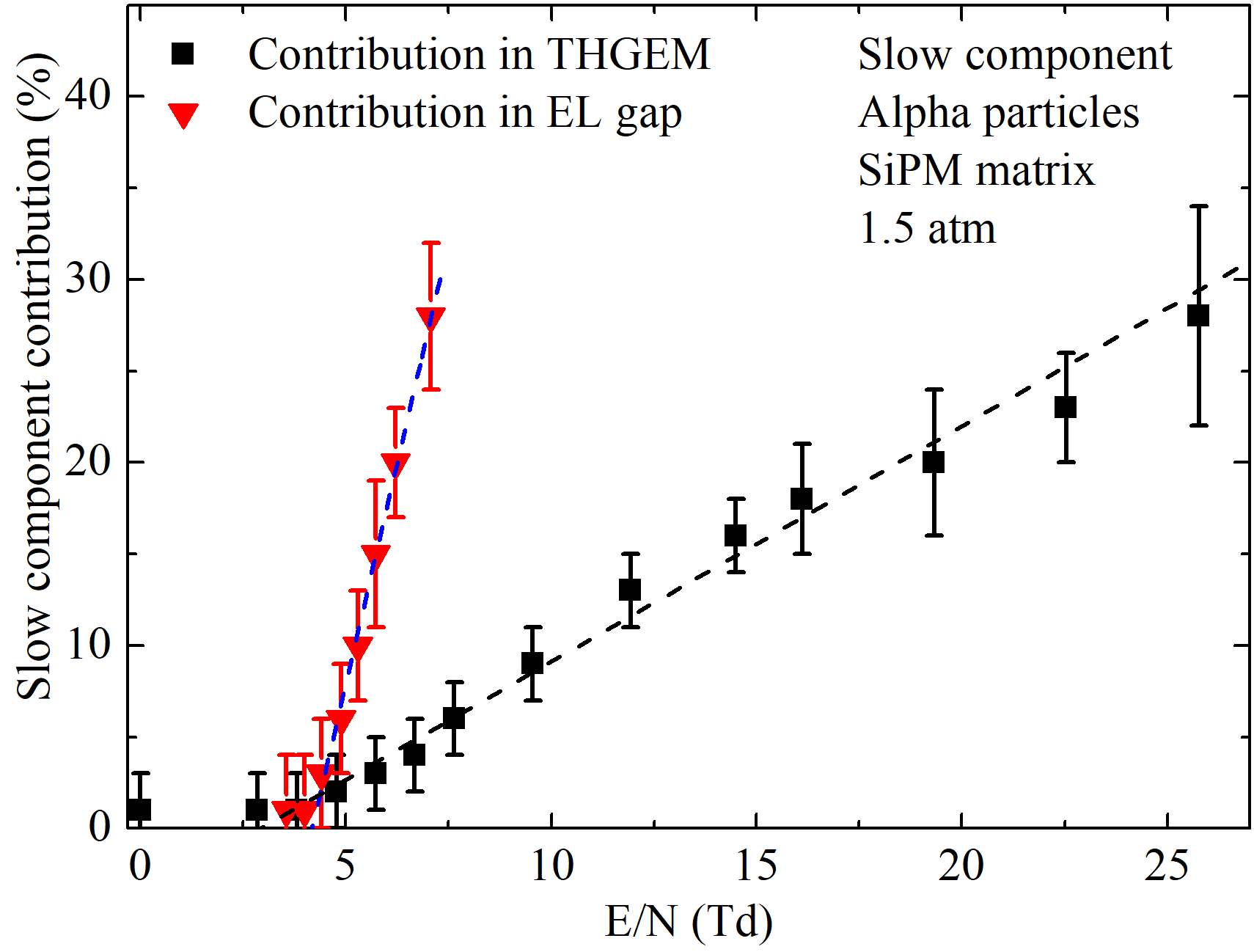}
\caption{Contribution to overall signal of the slow component of EL signal in THGEM and EL gap as a function of the reduced electric field in the THGEM hole center and EL gap respectively.
The data correspond to Fig.~\ref{fig:THGEM_vs_gap_tau} and were obtained together. The dashed lines are linear fits to the data points.
}
\vspace{-10pt}
\label{fig:THGEM_vs_gap_contr}
\end{figure}

Fig.~\ref{fig:THGEM_vs_gap_tau} and~\ref{fig:THGEM_vs_gap_contr} compare the unusual slow components produced in the EL gap and in THGEM by their time constants and contribution to the overall signal respectively. These results indicate that the unusual slow components in the EL gap and THGEM are of the same nature. First of all, the appearances of the slow components in THGEM and EL gap occur at the same thresholds in the reduced electric field. In addition, these components have similar characteristic increase with electric field described by a linear function.

From Fig.~\ref{fig:THGEM_vs_gap_tau} and~\ref{fig:THGEM_vs_gap_contr} it can also be seen that the unusual slow components produced in the EL gap and THGEM at a given electric field are slightly different in time constants, by about 20\%, and considerably different in contributions, by about a factor of 6. This is easily explained in our hypothesis~\cite{Buzulutskov22} that the unusual slow components are due to electron trapping in the gas phase on metastable negative ions. If Poisson statistics of electron trapping can be applied to describe slow component contribution, then from Fig.~\ref{fig:THGEM_vs_gap_contr} one can conclude that mean drift path of the electron (trapping length) is greater than the EL gap or THGEM effective thickness for all electric fields. Indeed, if mean drift path is equal to the effective gap thickness, then the contribution of both slow and long components is 63\%, whereas the overall contribution does not exceed 50\% in the experiment. However, it should be remarked that Poisson statistics can describe slow components contribution in EL signal only approximately, in particular due to the fact that electron which was trapped has still made contribution to the fast component and not only to the slow one. 

When the trapping length is larger than the gap thickness, the time constant is determined mostly by a single trapping event, i.e. by the negative ion lifetime, and to a much lesser extent by multiple trapping events, the probability of the latter being well below unity. This explains why the increase of time constant with electric field is so weak. The large difference in contributions observed in Fig.~\ref{fig:THGEM_vs_gap_contr} for the EL gap and THGEM at a given electric field, of about a factor of 6, is consistent with different trapping probabilities for effective gap thicknesses varying by more than an order of magnitude: 10 mm versus 0.4~mm respectively.

It must be remarked, however, that simple comparison of the EL gap and THGEM data using mean drift path of an electron before its trapping is valid only qualitatively but not quantitatively for several reasons. First of all, THGEM holes have non-uniform electric field which can leak from the holes and affect values of electric field on distances from THGEM of about half its thickness (see Fig.~\ref{fig:THGEM_field} and~\cite{Bondar19}). Detailed simulation of electron drift and trapping in the holes is required to account for this effect.

\begin{figure}[!t]
\centering
\includegraphics[width=0.7\linewidth]{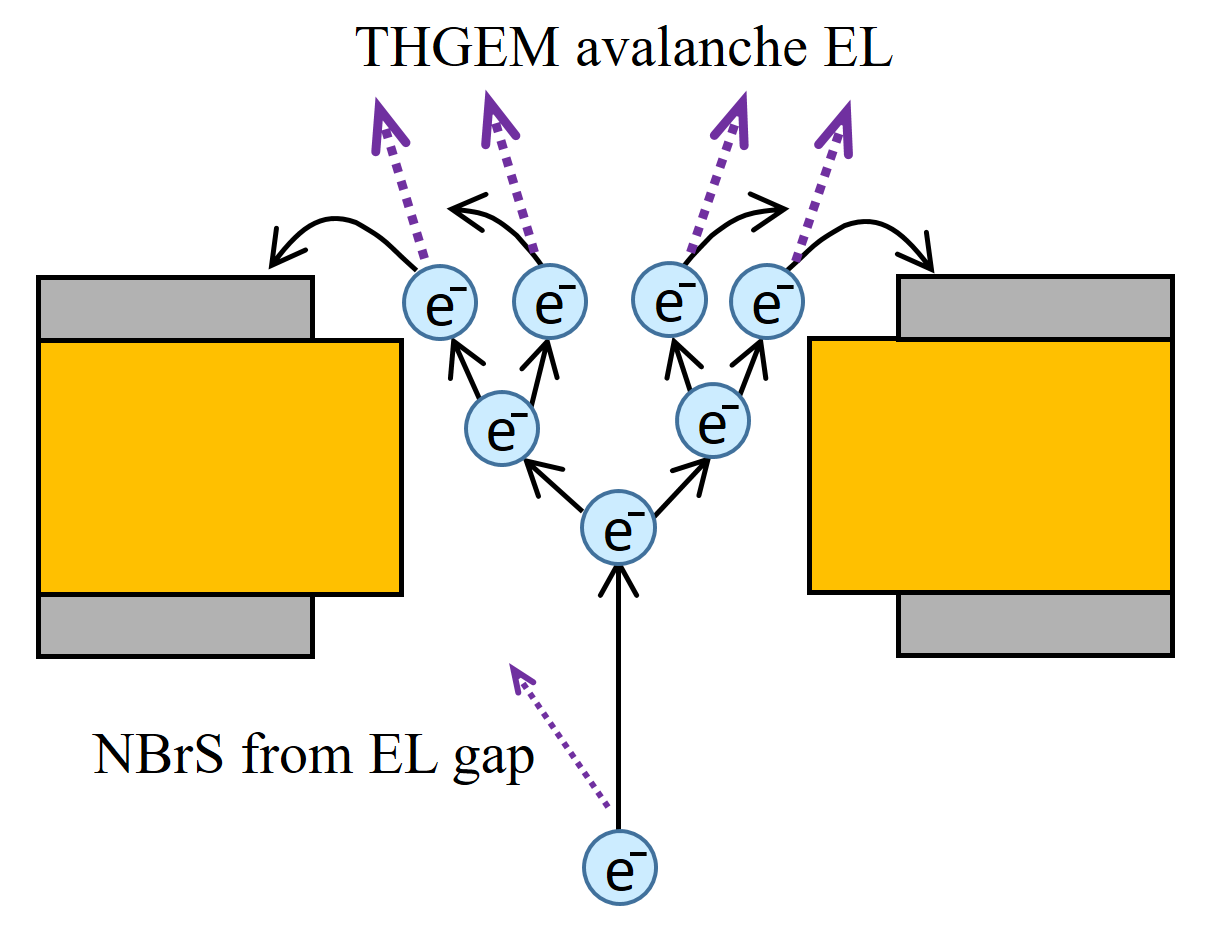}
\caption{Schematic view of electron avalanche occurring in THGEM hole at high voltages across THGEM. In this scenario, most of the EL signal is produced at the end of the avalanche.}
\vspace{-10pt}
\label{fig:THGEM_avalanche}
\end{figure}

Secondly, the observed values of slow component contribution to the EL signal do not correspond directly to probability of whether electron was trapped or not during its drift through whole distance. Indeed, if an electron was trapped in the middle of its drift, it still has contributed to the fast component, albeit with reduced weight. This biasing towards increasing fast component contribution means that Poisson statistics can only be applied for estimates that will have an uncertainty of a factor of about 2.

The situation is complicated further if THGEM works in avalanche multiplication mode. In this case, most of the EL signal is produced at the end of an avalanche, as illustrated in Fig.~\ref{fig:THGEM_avalanche}. Under this condition there is no biasing towards increasing fast component contribution. Hence pulse shape of EL signal from avalanche EL reflects time profile of charge arriving to the end of the avalanche. Such correspondence between avalanche EL and charge arrival time is possible due to atomic EL being fast, with $\tau$ = 20--40~ns~\cite{Buzulutskov17}. This correspondence is used in the following section to show that the unusual slow components produced in the EL gap are present in the charge signal. However, this effect also prevents direct comparison of data obtained in avalanche mode with data obtained in proportional EL mode.

All issues discussed above mean that precise model of slow component production is required for quantitative description of all measurements. The model and accompanying Monte-Carlo simulations are currently being developed in our laboratory.

The fact that the unusual slow components can be produced in THGEM independently from the EL gap proves that the production occurs in the gas phase and not in the liquid bulk or at the liquid-gas interface. This in turn means that the unusual slow components are not related to known electronegative impurities. 

Indeed, given the upper limit of 3~ppb of O$_2$ impurity in liquid Ar (see section~\ref{SetupSection}), one can calculate its content in gaseous Ar at 1.0~atm using Raoult’s law. Saturated vapor pressure for O$_2$ at 87~K is 0.71~atm~\cite{Richard91}, which yields its content of 2~ppb in the gas phase. Taking electron attachment rates to this impurity being proportional to the atomic density (worst case scenario), we obtain that electron lifetime in the gas phase is $>$28~ms. Given electron drift time of 2--4~$\mu$s, the impurities can hardly be responsible for the unusual slow components. In addition, the attachment rate of O$_2$ decreases with the electric field, contradicting with the opposite behavior of the unusual slow components. On the other hand, the effect of electron attachment to unknown impurity with boiling point below that of Ar, albeit much less probable, cannot be fully excluded.

The results of this section raise a question of what would happen to the slow component if the THGEM is replaced by a standard (thin) GEM~\cite{Sauli16}, the dielectric thickness of which is about an order of magnitude smaller (0.05 mm versus 0.4 mm). Assuming that the reduction factor of the slow component contribution when going from THGEM to GEM is about the same as when going from EL gap to THGEM, one would expect a substantial suppression of the slow component in GEM compared to THGEM. This means that the GEM signal remains fast in two-phase Ar detectors, in contrast to that of THGEM; and this has been indeed confirmed elsewhere~\cite{Buzulutskov20,Bondar08}.

\section{Direct measurements of the charge-signal pulse shape}\label{results_charge}

\begin{figure}[!t]
\centering
\includegraphics[width=1.0\linewidth]{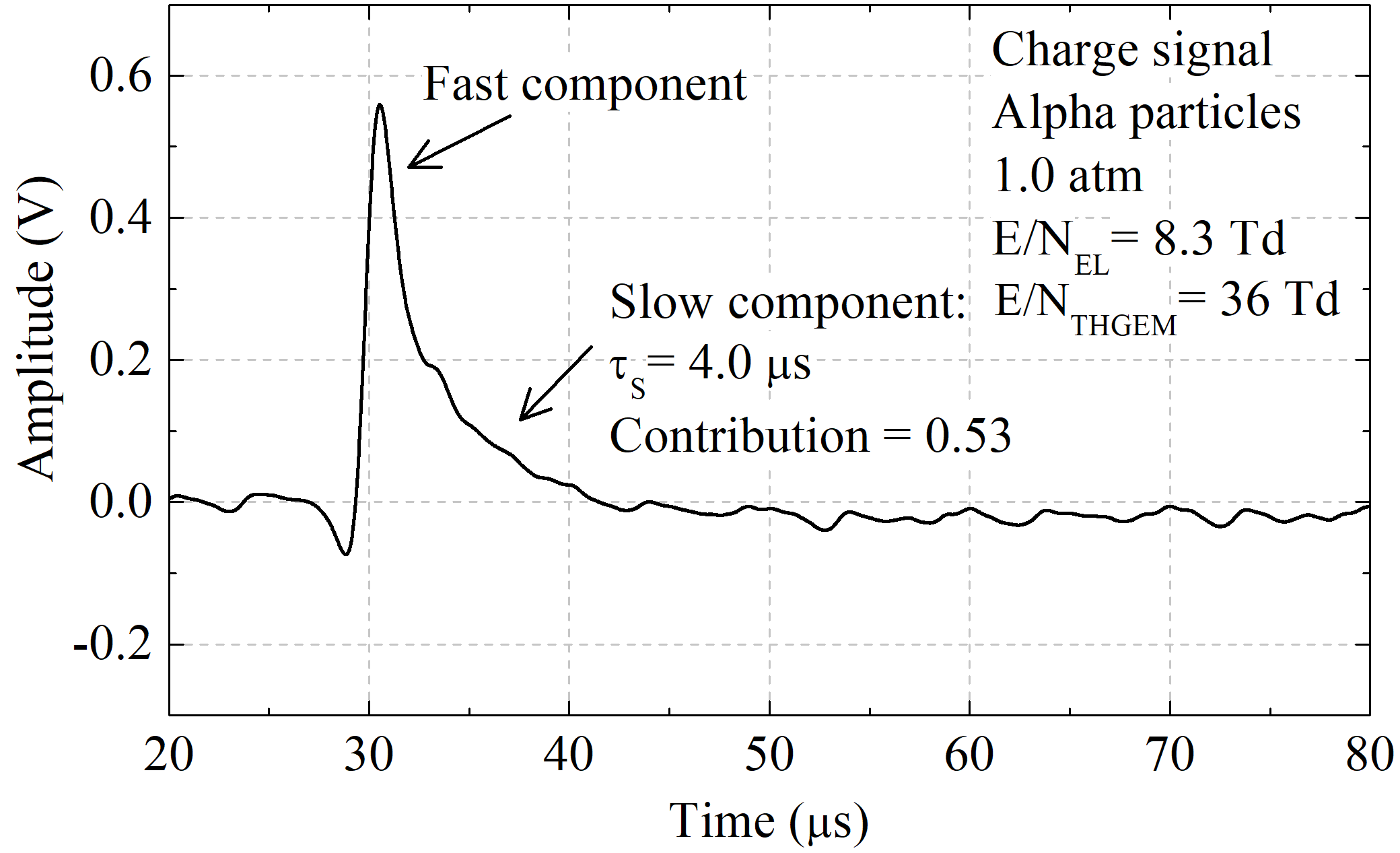}
\caption{Pulse shape of THGEM charge signal obtained in two-phase Ar detector at 1.0~atm, with $^{238}$Pu alpha particles, 10~mm thick EL gap, EL gap field of 8.3~Td and THGEM voltage of 2110~V, the latter corresponding to the field at THGEM hole center of 35.9~Td and effective THGEM charge gain of 20. The shaping time of charge readout circuit was 0.5~$\mu$s.}
\vspace{-10pt}
\label{fig:charge_shape}
\end{figure}

\begin{figure}[!t]
\centering
\includegraphics[width=1.0\linewidth]{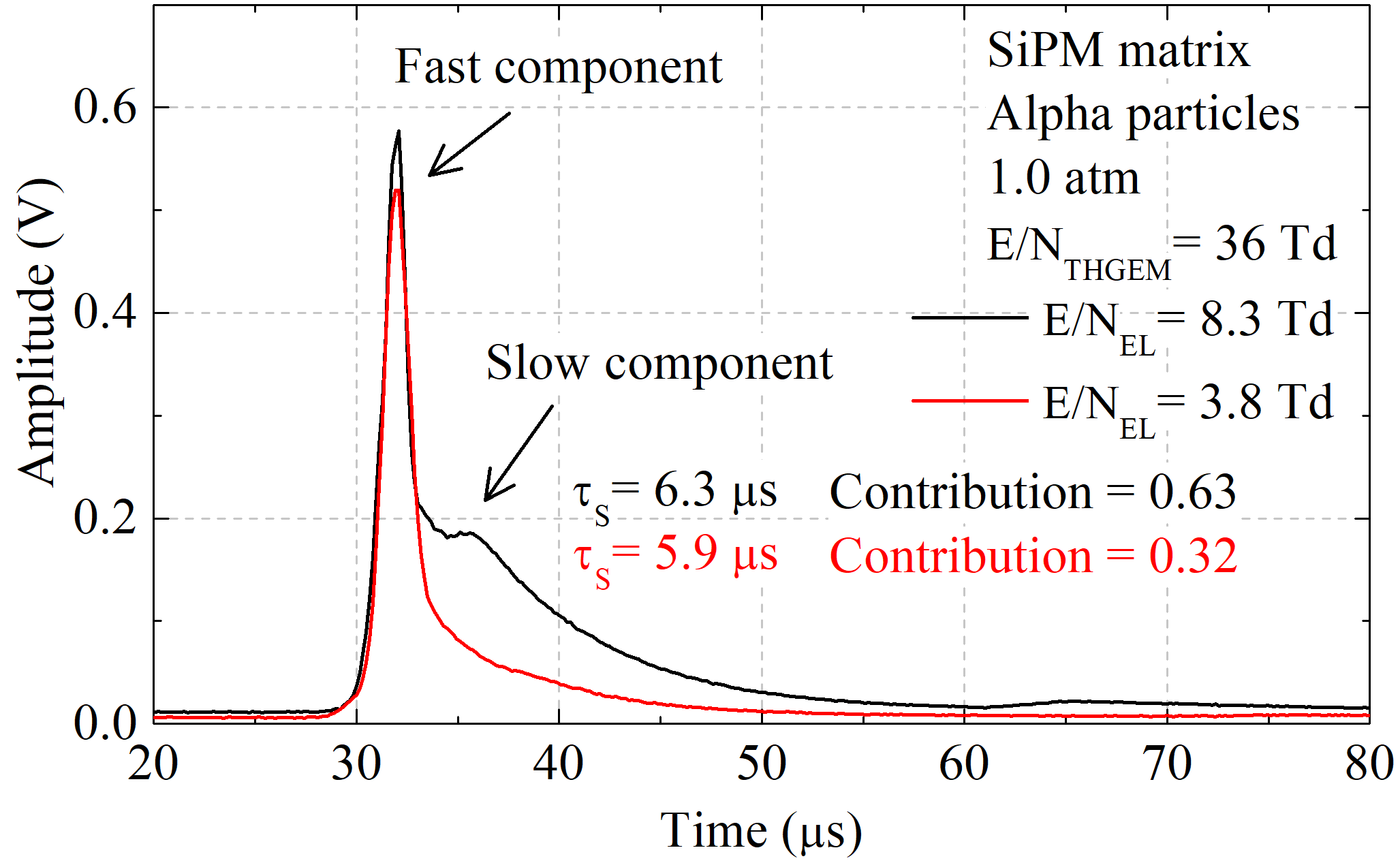}
\caption{Pulse shape of SiPM-matrix signal reflecting that of THGEM avalanche charge signal, at high (8.3~Td) and low (3.8~Td) EL gap field, obtained in two-phase Ar detector at 1.0~atm, with $^{238}$Pu alpha particles, 10~mm thick EL gap and THGEM voltage of 2110~V, the latter corresponding to the field at THGEM hole center of 35.9~Td and effective THGEM charge gain of 20.}
\vspace{-10pt}
\label{fig:SiPM_shape}
\end{figure}

To verify the hypothesis that the unusual slow components are produced due to delayed electrons temporarily trapped on metastable negative ions, the charge signal induced on the THGEM electrode by drifting electrons was directly measured. To maximize the signal-to-noise ratio, the maximal drift and EL gap fields, of 0.62~kV/cm and 8.3~Td respectively, were used. The THGEM voltage was 2110~V, corresponding to the effective THGEM charge gain of 20 at 1~atm (see Fig.~\ref{fig:THGEM_gain}). 


The measured charge-signal pulse shape is shown in Fig.~\ref{fig:charge_shape}. Here the time resolution was limited by a shaping time of the charge readout circuit of 0.5~$\mu$s. The pulse shape has a distinct fast and slow component, the latter corresponding to the unusual slow component observed before in EL signal, see Fig.~\ref{fig:S1_S2_shape} and \ref{fig:THGEM_shapes}. This result directly confirms the statement that the slow component in EL signal is produced in the charge signal itself, unambiguously confirming the effect of trapped electrons in S2 signal.

Compared to Fig.~\ref{fig:S1_S2_shape}, the transition from the fast to slow component is quite smooth due to poor time resolution.
The much better separation between the components is seen in Fig.~\ref{fig:SiPM_shape} where an optical signal of the SiPM-matrix was used instead of the THGEM charge signal, with superior time resolution. As discussed before in section~\ref{results_slow_comp} (also see Fig.~\ref{fig:THGEM_avalanche}), the SiPM-matrix optical signal directly reflects the avalanche charge of THGEM when it is operated in electron multiplication mode. It also means that the shape of fast component is defined by the dynamic of the avalanche occurring in THGEM hole, resulting in the shape of fast component being independent of the EL gap.

It should be noted that the bump seen at around 36~$\mu$s for black line in Fig.~\ref{fig:SiPM_shape} is much more pronounced than bumps seen in Fig.~\ref{fig:EL_gap_shapes_log} and~\ref{fig:THGEM_shapes_log}. It corresponds to also greater S3 signal (seen at around 65~$\mu$s in Fig.~\ref{fig:SiPM_shape}) which indicates that this bump is related to photo-ionization of the interface grid by EL photons as discussed in section~\ref{results_slow_comp}.

At lower EL gap fields, the slow component in Fig.~\ref{fig:SiPM_shape} is fully defined by that of THGEM itself (compare to Fig.~\ref{fig:THGEM_shapes}). On the other hand, at higher EL gap fields, the slow component is composed of both that of the EL gap and that of THGEM itself, with about the same contribution, as one can see from Fig.~\ref{fig:SiPM_shape}. Thus, it is confirmed that the slow component exists in both the charge signal of the EL gap and THGEM. Indeed, if the unusual slow component produced in the EL gap is due to some delayed scintillations or EL, it would not be seen against the background of large avalanche EL in THGEM. In this case, the pulse shape of avalanche EL would not depend on the conditions in the EL gap. 


\section{Conclusions}

In this work we explored the problem of slow components in two-phase Ar detectors for dark matter searches in a new way, namely using a Thick Gas Electron Multiplier (THGEM) coupled to the electroluminescence (EL) gap. There has been a hypothesis proposed in our previous work~\cite{Buzulutskov22}, that these slow components are produced by delayed electrons, temporarily trapped during their drift in the EL gap on metastable negative Ar ions. The nature of these ions is investigated in~\cite{Buzulutskov23c}.

This hypothesis is consistent with results of this work where an unusual slow component was observed in EL signal of THGEM, similar to that observed in the EL gap. Moreover, with the help of THGEM operated in an electron multiplication mode, the slow component was observed in the charge signal itself, confirming the effect of trapped electrons in S2 signal. On the other hand, the effect of electron attachment to unknown impurity, albeit much less probable, cannot be fully excluded.

\sloppy These trapped electrons are an additional background to spurious electrons observed in DarkSide-50 at 10~ms scales~\cite{Agnes18,Agnes23} which are associated with electron captures in the liquid bulk on electronegative impurities. Despite the fact that DarkSide's nominal reduced electric field is only 4.6~Td, below the threshold of emergence of unusual slow components, electrode sagging can increase the field at the center of the detector up to 5.6~Td~\cite{Bondar20a}. And while the unusual slow components do not affect the spurious electron background at time scales much greater than 50~$\mu$s, the effect may lead to misidentification of photoelectron cluster corresponding to single drifting electron. Indeed, due to electron trapping such cluster may be separated into several, leading to electron being not registered, depending on the analysis algorithm. Moreover, if electrons are trapped on metastable negative Ar ions in the gas phase, the similar process may occur in the liquid which would result in bulk spurious electron component independent of impurities.

Another conclusion of this work is that the signal of THGEM operated in electron multiplication mode in two-phase Ar is inherently slow due to the presence of slow component. This fact should be taken into account when planning the use of THGEM in two-phase Ar detectors. 

\section*{Acknowledgments}

This work was supported in part by Russian Science Foundation (project no. 20-12-00008, \url{https://rscf.ru/project/20-12-00008/}).

\bibliographystyle{spphys_modified}       
\bibliography{Manuscript}   

\end{document}